\documentclass[aps,prl,twocolumn,showpacs,amsmath,prepintnumbers,amssymb,superscriptaddress,reprint]{revtex4}
\usepackage{graphicx}
\usepackage{dcolumn}
\usepackage{color}
\usepackage[latin1]{inputenc}
\usepackage{comment}
\usepackage{amsmath}
\usepackage[normalem]{ulem}

\graphicspath{{}}
\begin{document}
%
\title{Coherent control over three-dimensional spin polarization for the spin-orbit coupled surface state of Bi$_{2}$Se$_{3}$}
\author{Kenta~Kuroda \footnote{These two authors contributed equally.}$^{,}$\footnote{kuroken224@issp.u-tokyo.ac.jp}}
\author{Koichiro~Yaji \footnotemark[1]}
\author{M.~Nakayama}
\author{A.~Harasawa}
\author{Y.~Ishida}
\affiliation{Institute for Solid State Physics (ISSP), University of Tokyo, Kashiwa, Chiba 277-8581, Japan}
\author{S.~Watanabe}
\affiliation{Research Institute for Science and Technology, Tokyo University of Science, Chiba, 277-8510, Japan}
\author{C.-T.~Chen}
\affiliation{Beijin Center for Crystal Research and Development, Chinese Academy of Science, Zhongguancun, Beijin 100190, China}
\author{T.~Kondo}
\author{F.~Komori \footnote{komori@issp.u-tokyo.ac.jp}}
\author{S.~Shin}
\affiliation{Institute for Solid State Physics (ISSP), University of Tokyo, Kashiwa, Chiba 277-8581, Japan}
\date{\today}
%
\begin{abstract}                                                   %
Interference of spin-up and spin-down eigenstates depicts spin rotation of electrons, which is a fundamental concept of quantum mechanics and accepts technological challenges for the electrical spin manipulation.
Here, we visualize this coherent spin physics through laser spin- and angle-resolved photoemission spectroscopy on a spin-orbital entangled surface-state of a topological insulator.
It is unambiguously revealed that the linearly polarized laser can simultaneously excite spin-up and spin-down states and these quantum spin-basis are coherently superposed in photoelectron states.
The superposition and the resulting spin rotation is arbitrary manipulated by the direction of the laser field.
Moreover, the full observation of the spin rotation displays the phase of the quantum states.
This presents a new facet of laser-photoemission technique for investigation of quantum spin physics opening new possibilities in the field of quantum spintronic applications.
\end{abstract}
\maketitle
%
%
%
Coherent manipulation of electron spin offers many potential applications in spintronic devices~\cite{Awschalom13Science} and spin-based quantum information science~\cite{Morton11Ann,Nowack07Science}.
The key is to take control over the superposition of spin-up and spin-down states resulting in interference. 
Spin-orbit coupling (SOC) mediates electric field and electron spin; thus electric/optical control of spins may become possible. 
Up to now, the coherent spin manipulation through electric fields is demonstrated in a few systems, such as quantum qubits~\cite{Nowack07Science, Hanson08Nature} and semiconductor-heterostructure interfaces~\cite{Kato03Nature}.
Optical control of spins utilizing the superimposed states in diamond is also demonstrated~\cite{Buckley10Science}.
In this Letter we introduce a great methodology which is capable of directly accessing this scheme in photoelectron spin, based on a combination of polarization-variable laser with spin- and angle-resolved photoemission spectroscopy (laser-SARPES).

We use the technique for a well-understood model system, a spin-polarized surface state of a topological insulator (TI), Bi$_{2}$Se$_{3}$. 
The TI has been discovered as a new class of matter, which is characterized by a metallic topological surface-state (TSS) intersecting the bulk band-gap~\cite{Hasan10rmp, Ando13jpsj}.
The TSS forms spin-polarized Dirac-cone-like energy dispersion~\cite{Xia09Nature, Chen09Science, Hsieh09Nature, Pan11prl, Souma11prl, Miyamoto12prl}.
In particular, as a consequence of the strong SOC, different spins and orbitals are mixed in the TSS wavefunction, which generates a spin-orbital entangled texture~\cite{Zhang13prl, Cao13NaturePhys, Zhu13prl}.
Since the electric field of light couples to the orbitals, polarized light can, in principle, selectively excite the fully spin-polarized electrons with either spin-up or spin-down from the spin-orbital entanglement.
The spin-selective excitation in the TSS by polarized light was theoretically studied~\cite{Park12PRL,Park16PRL} and demonstrated in previous SARPES experiments~\cite{Cao12arxiv, Zhu14prl, Xie14NatureComm, Joswiak13NaturePhys}.
The entangled spin-orbital texture of the TSS is therefore suitable for the coherent spin control as a novel source of spin-up and spin-down basis.
Indeed, Zhu $et\; al$ proposed a layer-dependent interference effect of the excited these two spin-basis and this was deduced from the photon energy and experimental configuration dependence~\cite{Zhu14prl}.
Yet, a clear experimental evidence of the coherent spin effect has not been achieved in the spin-orbital entangled texture.

In the present study, the interference effect between the two quantum spin basis is clearly demonstrated in systematic linear-polarization dependence of the photoelectron spin orientation by using laser-SARPES.
Although the polarization-sensitive photoelectron spin for the TSS is studied in previous SARPES experiments~\cite{Cao12arxiv, Zhu14prl, Xie14NatureComm, Joswiak13NaturePhys},
our methodology presents two novel advantages.
First, the superposition of the two quantum-spin basis and the resulting spin rotation-angle can be arbitrary controlled in three-dimension just by the direction of the laser field.
Second, the full observation of the coherent spin rotation moreover displays the relative phase of the quantum spin states.
This new methodology therefore dramatically increases the capabilities of conventional SARPES for directly accessing and visualizing the quantum-mechanically entangled states.

\begin{figure}[t!]
\begin{center}
\includegraphics[width=0.98\columnwidth]{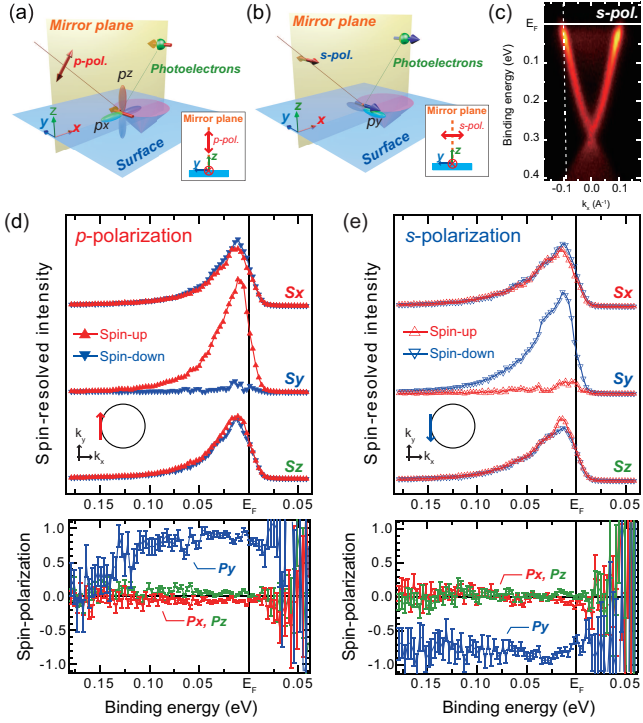}
\caption[]{
(a) and (b) Experimental configurations for $p$- and $s$-polarization where the light incidence plane and the detection plane match the mirror plane of the crystal ($x$-$z$ plane).
The $p$- ($s$-) polarization excites $even$ ($odd$) parity orbitals with respect to the mirror plane.
The $odd$ ($even$) orbital is derived from $p_{y}$ ($p_{z}$ and $p_{x}$) like orbital that is coupled to $+y$-spin ($-y$-spin) in the spin-orbital entangled texture~\cite{Zhang13prl, Cao13NaturePhys, Zhu13prl, Cao12arxiv, Zhu14prl, Xie14NatureComm}.
(c) ARPES intensity map along $\bar{\Gamma}$-$\bar{M}$ high symmetry line.
(d) and (e) Spin-resolved photoemission intensity and the corresponding spin polarization as a function of binding energy obtained by $p$- and $s$-polarizations, at a fixed emission angle corresponding to the dashed line-cut in (c).
The observed spin orientation is shown in the inset.}
\label{fig1}
\end{center}
\end{figure}
\begin{figure}[t]
\begin{center}
\includegraphics[width=0.95\columnwidth]{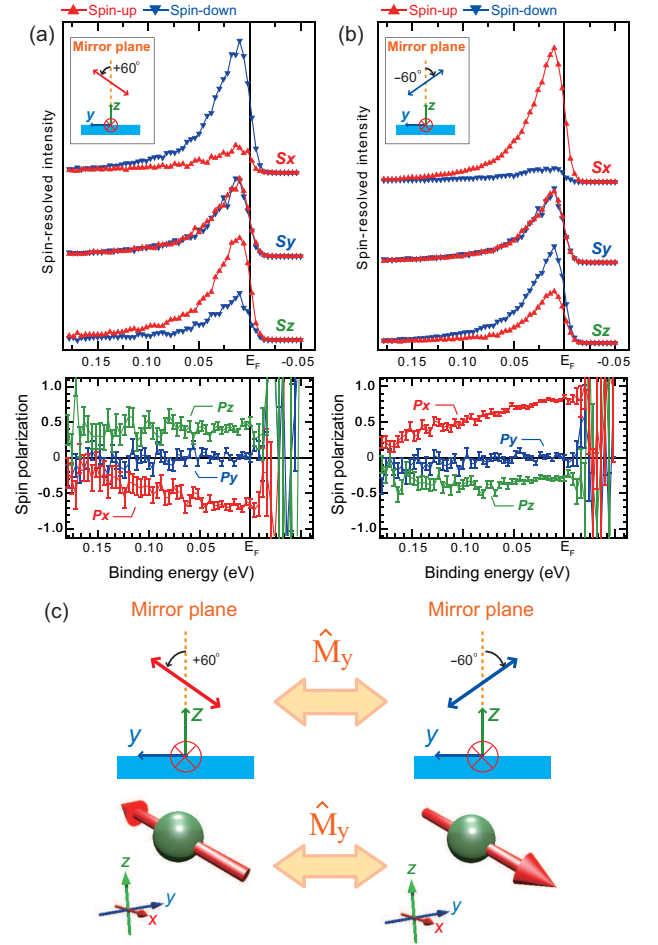}
\caption[]{(a) and (b) Three-dimensional spin-resolved photoemission intensity and the spin polarization as a function of binding energy at the same $k$ point as that shown in Fig.~\ref{fig1}(d) and (e). 
The insets show the linear polarization direction for the measurements.
These configurations are transformed to each other by the mirror symmetry operation ($\hat{M_{y}}$), which governs the orientation of the resulting spin as shown in (c).}
\label{fig2}
\end{center}
\end{figure}
%
Single crystalline samples of Bi$_{2}$Se$_{3}$ were grown by using the Bridgman method. 
Laser-SARPES experiments were performed at home-built laser-SARPES machine implemented by a high-flux 6.994-eV laser source~\cite{Yaji16rsi}.
The laser-SARPES machine is equipped with two high efficient VLEED (Very-Low Energy Electron Diffraction) spin detectors and a hemispherical analyzer with a photoelectron reflector function (SCIENTA-OMICRON DA30L).
This spectrometer resolves photoelectron spin-polarization in three dimension.
The experimental configuration are shown in Figs.~\ref{fig1}(a) and (b).
The photoelectrons were collected along $\bar{\Gamma}$-$\bar{M}$ high symmetry line (along $x$). 
During the measurement, the sample temperature was kept below 20~K, and instrumental energy and momentum resolutions were set below 20~meV and 0.7$^{\circ}$ for SARPES, respectively.
Clean surfaces were obtained by cleaving at low temperature $\sim$20~K under an ultra-high vacuum better than 1$\times$10$^{-8}$~Pa. 

%
\begin{figure*}[t]
\begin{center}
\includegraphics[width=0.8\textwidth]{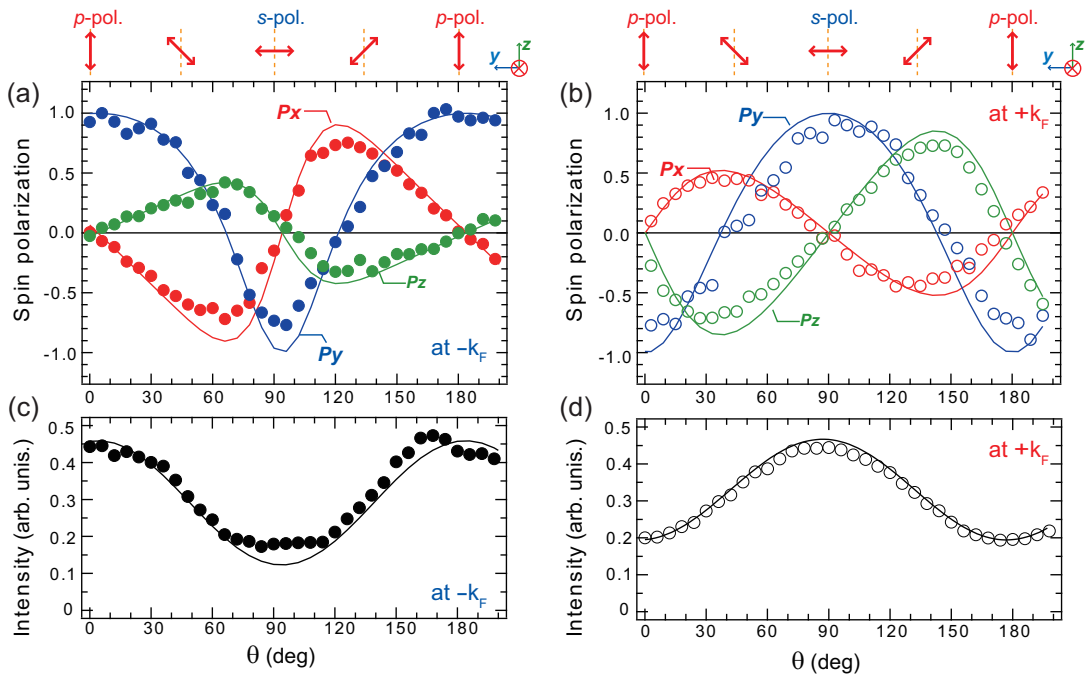}
\caption[]{(a) and (b) The plots of $P_{x,y,z}$ of the TSS as a function of the linear polarization angle ($\theta$) with respect to the mirror plane at $k_{x}$=$-k_{\rm{F}}$ and $+k_{\rm{F}}$, respectively. At $\theta$=0$^{\circ}$, the electric field of the laser is aligned in $x$-$z$ plane and thus $p$-polarization. From $\theta$=0$^{\circ}$ to 90$^{\circ}$, changes from $p$- to $s$-polarization, and becomes $p$-polarization again at $\theta$=180$^{\circ}$.
(c) and (d) The corresponding photoelectron intensity plots of the TSS as a function of the $\theta$.
The solid curves are obtained by fitting analysis with Eq.~(\ref{eq2}).
}
\label{fig3}
\end{center}
\end{figure*}
%
%
%
%
%
We start by describing the spin-resolved data for two different polarization geometries of $p$-polarization ($\epsilon_{p}$) and $s$-polarization ($\epsilon_{s}$) [Figs.~\ref{fig1}(a) and (b)].
Figure~\ref{fig1}(c) represents a standard ARPES intensity map.
The map shows a sharp Dirac-cone-like energy dispersion of the TSS.
Figures~\ref{fig1}(d) and (e) respectively show spin-resolved energy distribution curves with the corresponding spin polarization obtained at ($k_{x}$, $k_{y}$)=($-k_{\rm{F}}$, 0) [dashed line in Fig.~\ref{fig1}(c)] by using $\epsilon_{p}$ and $\epsilon_{s}$.
For the $y$-spin polarization ($P_{y}$), the result clearly shows spin selective excitation of the linearly polarized laser.
Since $P_{y}$ is fully polarized, the other components, $P_{x}$ and $P_{z}$, are negligibly small.
The observed fully spin-polarized photoelectrons and the spin selective excitation are in good agreement with the previous SARPES works~\cite{Joswiak13NaturePhys,Zhu14prl} and dipole selection rule for the spin-orbital texture [the figure caption of Fig.~\ref{fig1}].

The spin-polarized electrons with $\pm{y}$-spin can be simultaneously excited if the light polarization $\epsilon$ is tuned between $\epsilon_{p}$ and $\epsilon_{s}$, where one would expect a reduction of $P_{y}$ because the positive and negative $P_{y}$ excited by $\epsilon_{p}$ and $\epsilon_{s}$ components will cancel out.
Indeed the reduction of $|P_{y}|$ has been previously reported by rotating the linear polarization of the light~\cite{Park16PRL,Joswiak13NaturePhys}.
As is clearly seen in Figs.~\ref{fig2}(a) and (b), we also observe that $|P_{y}|$ significantly reduces to 0$\%$ when the linear polarization is rotated by $\theta$=$\pm$60$^{\circ}$ ($\epsilon_{+60}$ and $\epsilon_{-60}$) with respect to the mirror plane [see the inset figures].
The detection of the spin in all dimensions reveals the appearance of significant $P_{x,z}$ upon the deduction of $P_{y}$.
In particular, $|P_{x}|$ develops up to $\sim$80$\%$.

The observation of the finite $P_{x,z}$ may, at first glance, be surprising because the spin-orbital texture locks the spin of the TSS electrons into $y$ at ($-k_{\rm{F}}$, 0)~\cite{Zhang13prl, Cao13NaturePhys}.
Thus, the spin polarization effect in the photoelectrons cannot be explained by the spin-orbital textures alone but by photoemission matrix element effect~\cite{Zhu14prl,Ishida11prl,Heinzmann2012spin,Krasovskii15spin}.
In the following, we show that the symmetry of the experimental configuration plays a significant role in the resulting $P_{x,y,z}$.
We here define $P_{x,y,z}$ taken with ${\epsilon}$ at $k_{||}$ as $P_{x,y,z}$ (${\epsilon}$, $k_{x}$, $k_{y}$).
Since the mirror operation ($\hat{M_{y}}$) leaves the experimental set-up invariant for ${\epsilon}_{p}$ and ${\epsilon}_{s}$ [Figs.~\ref{fig1}(a) and (b)], the expectation value of $P_{x,y,z}$ has to be equivalent to those after $\hat{M_{y}}$.
One can therefore confirm $P_{x,z}$ ($\epsilon_{p,s}$, $k_x$, 0)=0, since $\hat{M_{y}}$ reverses the sign of $P_{x,z}$ but not that of $P_{y}$~\cite{SI}.
Thus, the symmetric experimental configuration for $\epsilon_{p}$ and $\epsilon_{s}$ prohibits finite $P_{x,y}$.
In contrast, the tilted electric field of $\epsilon_{+60}$ and $\epsilon_{-60}$ breaks the symmetry in the overall experimental configuration, which therefore allows non-vanishing $P_{x,z}$.
We note that $\epsilon_{+60}$ and $\epsilon_{-60}$ are antisymmetric with respect to $\hat{M_{y}}$ as schematically shown in Fig.~\ref{fig2}(c).
Correspondingly the resulting $P_{x,z}$ show antisymmetric spin polarization [Figs.~\ref{fig2}(a) and (b)].
Therefore, the spin orientation of the photoelectrons sensitively relies on the direction of the applied $\epsilon$ of the laser. 
\begin{figure}[t]
\begin{center}
\includegraphics[width=0.95\columnwidth]{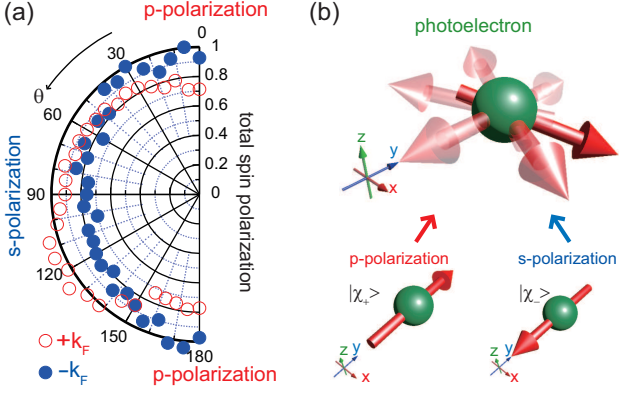}
\caption[]{(a) The polar plots of magnitude of the resulting total $P$ for (closed circles) $-k_{\rm{F}}$ and (open circles) $+k_{\rm{F}}$ as a function of the $\theta$.
(b) Summary of the mechanism for three-dimensional spin rotation due the interference of spin-up and spin-down $y$-spinors from the spin-orbital texture.
}
\label{fig4}
\end{center}
\end{figure}

Figure~\ref{fig3} represents the linear polarization evolution of $P_{x,y,z}$ [see also the supplemental movie S1].
In the data for $k_{x}$=$-k_{\rm{F}}$ [Fig.~\ref{fig3}(a)], $P_{y}$ decreases from +100$\%$ to $-$100$\%$ by rotating linear polarization from $\epsilon_{p}$ ($\theta$=0$^{\circ}$) to $\epsilon_{s}$ ($\theta$=90$^{\circ}$) and it vanishes at the specific $\theta$  ($\theta_{0}$) $\sim$60$^{\circ}$ and 120$^{\circ}$.
The data shows that $|P_{x,z}|$ achieves the maximum when $P_{y}$ is vanished.
Due to the mirror symmetry as we described above, $P_{x,z}$ ($P_{y}$) is antisymmetric (symmetric) with respect to that at $\theta$=90$^{\circ}$.
Although the data for $k_{x}$=$+k_{\rm{F}}$ exhibits the similar linear-polarization dependence, the sign of $P_{x,y,z}$ is reversed from that for $k_{x}$=$-k_{\rm{F}}$, and the $\theta_{0}$ clearly shifts to $\sim$30$^{\circ}$ and 150$^{\circ}$ [Fig.~\ref{fig3}(b)].

In order to describe the $\epsilon$ dependence of $P_{x,y, z}$, we built a model for $P_{x,y,z}$ by taking account of the spin-orbital entanglement of the TSS and the photoemission matrix element.
We write the TSS wavefunction at ($-k_{\rm{F}}$, 0)~\cite{Zhang13prl,Cao13NaturePhys} as follows:
\begin{eqnarray}
|\psi_{\rm{initial}}>=w_{\uparrow_{y}}|{\rm{even}}, \chi_{+}>+w_{\downarrow_{y}}|{\rm{odd}}, \chi_{-}>.
\label{eq1}
\end{eqnarray}
Here, $w_{{\uparrow}y}$ ($w_{{\downarrow}y}$) is a complex coefficient of an individual $even$ ($odd$) parity orbital component coupled to a spinor  $\chi_{+}$ ($\chi_{-}$) quantized along $y$, and $w_{{\uparrow}y ({\downarrow}y)}$ consists of a combination of real and imaginary parts, $w^{r}_{{\uparrow}y ({\downarrow}y)}$+$iw^{i}_{{\uparrow}y ({\downarrow}y)}$.
When $\epsilon$ is rotated in between $\epsilon_{p}$ and $\epsilon_{s}$, these polarization components simultaneously excite the spin-up and spin-down states into a particular photoelectron state.
Our model treats $P_{x,z}$ as a consequence of the coherent superposition of the two spin basis along $y$ rather than an incoherent sum of photoelectrons with $\pm{P_{y}}$.
We obtain the expectation values of the spin polarization vector as follows~\cite{SI}:
\begin{eqnarray}
&P_{x}&=i\frac{w_{\uparrow_{y}}w^{\ast}_{\downarrow_{y}}m_{\rm{even}}m^{\ast}_{\rm{odd}}
-w_{\downarrow_{y}}w^{\ast}_{\uparrow_{y}}m_{\rm{odd}}m^{\ast}_{\rm{even}}}
{I_{\rm{total}}}{\rm{cos}}\theta{\rm{sin}}\theta\nonumber ,\\
&P_{y}&=\frac{|w_{\uparrow_{y}}m_{\rm{even}}{\rm{cos}}\theta|^{2}-|w_{\downarrow_{y}}m_{\rm{odd}}{\rm{sin}}\theta|^{2}}
{I_{\rm{total}}}\nonumber ,\\
&P_{z}&=\frac{w_{\uparrow_{y}}w^{\ast}_{\downarrow_{y}}m_{\rm{even}}m^{\ast}_{\rm{odd}}
+w_{\downarrow_{y}}w^{\ast}_{\uparrow_{y}}m_{\rm{odd}}m^{\ast}_{\rm{even}}}
{I_{\rm{total}}}{\rm{cos}}\theta{\rm{sin}}\theta\nonumber ,\\
\label{eq2}
\end{eqnarray}
where $m_{even}$ ($m_{odd}$) is the dipole matrix element for $\epsilon_{p}$ ($\epsilon_{s}$), respectively.
Equation~(\ref{eq2}) explains that the photoelectron $P_{x,z}$ can be non-zero if $\epsilon$ is between $\epsilon_{p}$ and $\epsilon_{s}$.
In addition, Eq.~(\ref{eq2}) describes the specific angle $\theta_{0}$ where $P_{y}$ vanishes as:
\begin{eqnarray}
\rm{tan^{-1}\theta_{0}}=\it{\frac{|w_{\uparrow_{y}}m_{even}|}{|w_{\downarrow_{y}}m_{odd}|}=\sqrt{\frac{I_{\epsilon_{p}}}{I_{\epsilon_{s}}}}}.
\label{eq3}
\end{eqnarray}
One can learn that $\theta_{0}$ depends on the photoelectron-intensity ratio for $\epsilon_{p}$ ($I_{\epsilon_{p}}$) and $\epsilon_{s}$ ($I_{\epsilon_{s}}$).
Equations~(\ref{eq2}) and (\ref{eq3}) result from the dipole selection rule collaborated with the spin-orbital entanglement of Eq.~(\ref{eq1}).
%

The model reproduces the light-polarization dependence and the position of $\theta_{0}$ for $k_{x}$=${\pm}k_{\rm{F}}$ [solid lines in Figs.~\ref{fig3}(a) and (b)].
The relationship between $I_{\epsilon_{p}}$ and $I_{\epsilon_{s}}$ for $k_{x}$=$+k_{\rm{F}}$ reverses in the data for $k_{x}$=$-k_{\rm{F}}$: $I_{\epsilon_{p}}$ is higher than $I_{\epsilon_{s}}$ at ${\pm}k_{\rm{F}}$ [Fig.~\ref{fig3}(c)] but $I_{\epsilon_{p}}$ is lower than $I_{\epsilon_{s}}$ at $+k_{\rm{F}}$ [Fig.~\ref{fig3}(d)].
Equation~(\ref{eq3}) explains that this causes the deviation of $\theta_{0}$ between ${\pm}k_{\rm{F}}$ and the different shape of the polarization dependence. 
The coherent spin rotation via interference has to conserve the magnitude of the total spin, $\sqrt{|P_{x}|^{2}+|P_{y}|^{2}+|P_{z}|^{2}}$=1, for any linear polarization.
Figure~\ref{fig4}(a) presents $|P|$ as a function of the linear polarization angle.
Even though $\epsilon$ rotates the spin from original spin texture of the TSS, the spin is conserved, which also shows good agreement with previous works~\cite{Zhu14prl, Park12PRL}.
Therefore, the overall consistency between the model and the experimental results demonstrates the coherent rotation of the photoelectron spin by turning the light polarization. 

Since the coherent spin rotation sensitively relies on the phase of the quantum-mechanical states, the full observation of the photoelectron spin with our laser-SARPES extracts the phase information by the fitting analysis with Eq.~(\ref{eq2}).
If the fitting analysis is simplified by assuming $m_{even}$=$m_{odd}$, one can obtain the complex coefficient of Eq.~(\ref{eq1}), ($w^{r}_{\uparrow,y}$, $w^{i}_{\uparrow,y}$)=(0.6, 0.7) and ($w^{r}_{\downarrow,y}$, $w^{i}_{\downarrow,y}$)=(0.4, $-$0.1) at $-k_{\rm{F}}$.
The initial spin polarization in the TSS is thus derived to be +70$\%$, which is comparable to the theoretical prediction~\cite{Zhang13prl, Yazyev10prl}.
Although this interpretation could be much simplified and has to be compared to photoemission calculation with more correctly considering photoemission processes~\cite{Heinzmann2012spin, Krasovskii15spin, Barriga14prx} and layer-by-layer texture previously proposed~\cite{Zhu13prl}, this quantum phase-sensitive measurement presents great advantages of achieving coherent spin rotation with laser-SARPES for the investigation of quantum spin physics. 

%
In conclusion, we have shown the coherent spin effect in laser-SARPES that is involved with the optical selection rule and the spin-orbital entanglement of the TSS.
The mechanism is summarized in Fig.~\ref{fig4}(b).
If the linear polarization simultaneously excites spin-up and spin-down states, these two quantum spin-basis are coherently superposed in photoelectron state resulting in the spin rotation.
The superposition and the resulting spin rotation is arbitrary controlled by the direction of the laser field.
Moreover, in contrast to the conventional SARPES that merely relies on the spin-resolved photoelectron intensity, the spin rotation by using our laser-SARPES displays the phase of the quantum states.
Our results therefore dramatically increases the capabilities of SARPES to directly access and investigate the spin-dependent quantum interference.

%
We gratefully acknowledge funding JSPS Grant-in-Aid for Scientific Research (B) through Projects No. 26287061 and for Young Scientists (B) through Projects No. 15K17675.
%

\bibliographystyle{prsty}

\end{document}